# Hold and Feel! A Multiplayer Video Game System with Interpersonal Vibrotactile Feedback via Bracelet Controllers

Kenta Ebina and Taku Hachisu

*University of Tsukuba, Ibaraki, Japan*

(Email: {ebina, hachisu}@ah.iit.tsukuba.ac.jp)

**Abstract ---** Recent multiplayer video games (MPVGs) have increasingly incorporated social behaviors, such as gathering in the same place and facing each other, to enhance player interaction. This study aims to develop an MPVG system that facilitates social interaction through interpersonal touch between players. We present an MPVG system where two players wearing bracelet-type game controllers control a player character through touch to catch items and feel the position of the player character through vibrotactile sensations between their hands.

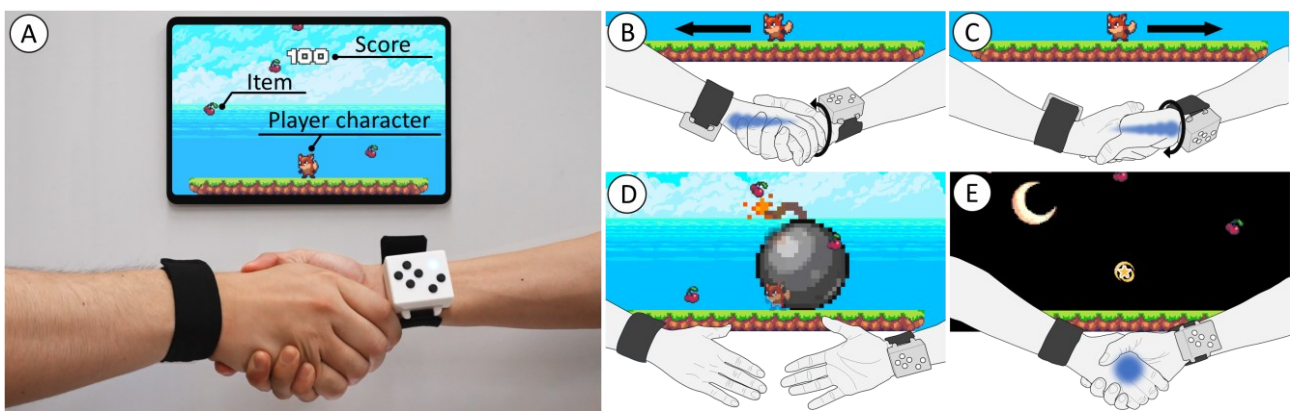

Fig.1 MPVG system: A) Controlling the player character (fox) through interpersonal touch and scoring points by catching items; B) Moving the fox to the left; C) Moving the fox to the right; D) Avoiding bombs by releasing the hands; E) Using vibrotactile cues to control the fox during the night scene.

## 1 INTRODUCTION

Multiplayer video games (MPVGs) facilitate social interaction by allowing multiple players to compete or collaborate within the game. Recently, there has been a growing trend toward incorporating social behaviors into gameplay, not just through controllers and monitors. This trend includes examples such as Niantic's Pokémon Go, which encourages players to gather in physical locations, and Nintendo's 1-2-Switch, which facilitates direct face-to-face interactions.

Interpersonal touch is a social behavior characterized by minimal social and physical distance, known to facilitate prosocial behavior [1]. Therefore, it holds promise as a means to enhance social interaction among players in MPVGs. In human-computer interaction research, MPVGs that use interpersonal touch between multiple players as input have been developed [2][3]. A key technical challenge in such studies is establishing technology that does not interfere with natural interpersonal touch due to bulky setups.

This study aims to develop an MPVG system that facilitates social interaction by utilizing interpersonal touch between players. We previously developed EnhancedTouch, a set of wireless bracelet devices that measure interpersonal skin-to-skin contact and provide real-time vibrotactile feedback [4]. This demonstration presents an MPVG system in which two players wearing bracelet-type game controllers move the player character (a fox) through touch to catch items (cherries) and feel the position of the vibration through vibrotactile sensations between their hands (Fig.1A-C and E). To earn higher points, this action game prompts players to synchronize their gestures and pay attention to the vibrotactile sensations when visual feedback is not available from the monitor (Fig.1E).

## 2 SYSTEM

The system consists of two bracelet-type game controllers (a transmitter and a receiver) and a tablet computer. Each bracelet comprises control boards, a 3D-printed case, and a wristband. On the top of the case, a D-pad and two buttons are mounted (Fig.1). The control board includes a microcontroller (Seeed Studio, XIAO ESP32C3), full-color LED (Cree LED, CLP6C-FKB-CK1P1G1BB7R3R3), linear resonant actuator (LRA) driver (Texas Instruments, DRV2605), LRA (ALPS ALPINE, HAPTIC Reactor Hybrid Tough Type), inertia sensor (TDK InvenSense, MPU-6050), custom inter-body area network (interBAN) module, a lithium-ion polymer battery, and two electrodes.

The controllers utilize interBAN technology [5] to measure interpersonal touches between two players. The transmitter continuously outputs a 3.3-$V_{PP}$, 10.7-MHz square wave signal to the wrist skin through one of the electrodes on the bottom surface of the case. When the players' hands come into contact, the signal is transmitted through the hands to the receiver's electrode. The signal strength varies primarily based on the contact area of the hands, allowing the microcontroller to measure the signal strength via its analog-to-digital converter and detect three states: no touch (no signal); gentle touch (weak signal); and strong touch (strong signal).

The microcontroller sends the button press states, inertia sensor values, and, for the receiver only, the interBAN signal strength to the computer via BLE, which are used as inputs to the game. Additionally, it receives command values from the computer to drive the LRA and LED as game outputs.

The computer executes a game application developed using the Unity game engine. Besides providing video output on the monitor, the computer calculates command values for the two bracelets. Our vibrotactile feedback technique [6] is implemented in the algorithm for calculating LRA command values. This technique creates an illusory vibrotactile sensation (phantom sensation [7]) localized between the hands, based on the intensity ratio of the two devices.

## 3 MULTIPLAYER VIDEO GAME

The goal of the game is to achieve a higher score by controlling the fox to catch as many falling cherries as possible (Fig.1), while avoiding falling bombs. Each gameplay session lasts 2 minutes. Participants are first instructed to wear the controllers. A demonstrator then explains how to control the fox's movement. To move the fox, the pair needs to hold each other's hands, which makes the fox opaque (controllable state). During this state, the pair can move the fox horizontally by performing hand supination/pronation (Fig.1B and C). When the pair releases their hands, the fox becomes semi-transparent (uncontrollable state). During this state, the pair cannot move the fox, and the fox loses collision detection with items. However, the fox can still avoid bombs (Fig.1D). This game design encourages synchronization of gestures, promoting cooperation.

The fox's position on the stage is represented not only on the monitor but also through vibrotactile sensations between the hands (Fig. 1B-C and E). A short vibration feedback is provided for each step the fox takes. During the night scene, when players cannot see the fox's position on the monitor, they must rely on this vibrotactile cue to control the fox by focusing on the sensations between their hands (Fig.1E).